\documentclass[aps,prl,nobibnotes,superscriptaddress,twocolumn,floatfix,amsfonts,amssymb,amsmath]{revtex4-1}
\usepackage{mhchem,bm,graphicx,float}
\usepackage[separate-uncertainty = true,per-mode=symbol]{siunitx}
\usepackage{natbib} 

\begin{document}
	
\title{Electronic Compressibility of Magic-Angle Graphene Superlattices}

\author{S.L. Tomarken}
\affiliation{Department of Physics, Massachusetts Institute of Technology, Cambridge, MA 02139}

\author{Y. Cao}
\affiliation{Department of Physics, Massachusetts Institute of Technology, Cambridge, MA 02139}

\author{A. Demir}
\affiliation{Department of Physics, Massachusetts Institute of Technology, Cambridge, MA 02139}

\author{K. Watanabe}
\affiliation{National Institute of Materials Science 1-1 Namiki, Tsukuba 305-0044, Japan}

\author{T. Taniguchi}
\affiliation{National Institute of Materials Science 1-1 Namiki, Tsukuba 305-0044, Japan}

\author{P. Jarillo-Herrero}
\email{Correspondence to: pjarillo@mit.edu}
\affiliation{Department of Physics, Massachusetts Institute of Technology, Cambridge, MA 02139}

\author{R.C. Ashoori}
\email{Correspondence to: ashoori@mit.edu}
\affiliation{Department of Physics, Massachusetts Institute of Technology, Cambridge, MA 02139}
 
\begin{abstract}
We report the first electronic compressibility measurements of magic-angle twisted bilayer graphene. The evolution of the compressibility with carrier density offers insights into the interaction-driven ground state that have not been accessible in prior transport and tunneling studies. From capacitance measurements, we determine chemical potential as a function of carrier density and find the widths of the energy gaps at fractional filling of the moir\'{e} lattice. In the electron-doped regime, we observe unexpectedly large gaps at quarter- and half-filling and strong electron--hole asymmetry. Moreover, we measure a $\mathord{\sim}\SI{35}{\milli\electronvolt}$ mini-bandwidth that is much wider than most theoretical estimates. Finally, we explore the field dependence up to the quantum Hall regime and observe significant differences from transport measurements.  
\end{abstract}

\maketitle
In most metals, interactions between electrons are sufficiently weak compared with electronic kinetic energy that they can be considered as a perturbation when calculating band structure. Because the kinetic energy of an electronic system is proportional to the bandwidth of its low energy bands, one route to finding materials with effectively strong electron--electron interactions is to study systems with flat energy dispersions. Stacking and rotating two monolayers of graphene by a controllable angle between the two layers can tune the resulting low-energy band structure through a large range of energy dispersion \cite{Morell-2010,Trambly2010-ui,Bistritzer2011-fm}. At certain small twist angles known as {\em magic angles}, the interlayer hybridization energy concentrates the low-energy density of states within about $\SI{10}{\milli\electronvolt}$ according to calculations \cite{Morell-2010,Bistritzer2011-fm}, providing a highly tunable test bed for strongly correlated physics. 

Early scanning tunneling spectroscopy (STS) measurements on twisted bilayer graphene revealed van Hove singularities in the low-energy moir\'{e} bands for low twist angle samples \cite{Li-2010}. More recent experiments show anomalous insulating behavior in magic-angle twisted bilayer graphene (MATBG) around $n=\pm n_\textrm{s}/2$, where $n_\textrm{s}$ is the electron density at which the low-energy superlattice bands are completely filled (four electrons per moir\'{e} cell). The insulating behavior has features similar to those expected for a Mott insulator arising from on-site Coulomb repulsion \cite{Cao2018-dj}, though there are currently many theoretical proposals for various strongly correlated phases \cite{Irkhin2018-mj,Padhi2018-un,Ochi2018-ap,Thomson2018-pt,Dodaro2018-zb,Xu2018-vg,Koshino-2018,Vafek-2019,Po2018,Liu2018-dw,Isobe2018-jv,Tarnopolsky-2019,Padhi-2018-2,Yuan-2019-vh}. Strikingly, superconductivity was also observed around $n=-n_\textrm{s}/2$ (two holes per supercell) with density-dependent critical temperatures up to \SI{1.7}{\kelvin} \cite{Cao2018-bf}.

\begin{figure}[h!tp]
\centering
\includegraphics[scale=1]{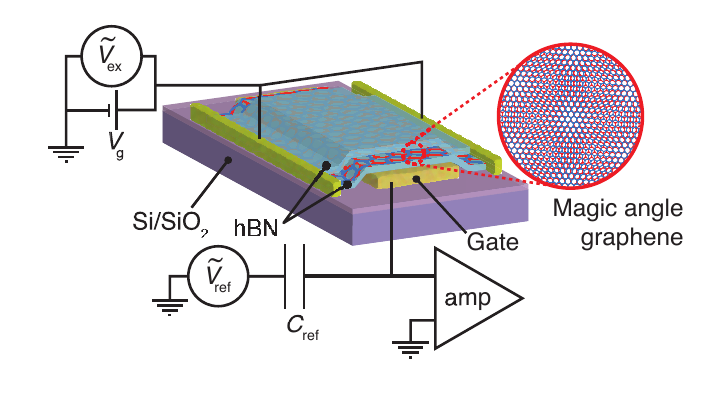}
\caption{Schematic of MATBG device and capacitance bridge measurement scheme.}
\label{schematic}
\end{figure}

In order to probe the thermodynamic ground state of MATBG, we use a low-temperature capacitance bridge to access the electronic compressibility of the two devices originally characterized with transport in Ref.~\onlinecite{Cao2018-bf}. By measuring the compressibility as a function of carrier density, we study the thermodynamic evolution of the interaction-driven phases at fractional filling, offering new insights into the nature of the MATBG ground state that complement previous transport \cite{Cao2018-dj,Cao2018-bf,Yankowitz2018-qv,Sharpe-2019,Polshyn-2019,Efetov-2019} and tunneling efforts \cite{Kerelsky2018-py,Choi-2019}. Additionally, we can extract the thermodynamic bandwidth of the low energy bands, enabling us to evaluate their relative flatness and compare to electronic structure calculations. Figure~\ref{schematic} shows the device geometry and measurement schematic. MATBG samples were fabricated using a ``tear-and-stack'' technique \cite{Kim-2016-rot, Cao2016-ld}. The MATBG is encapsulated between two layers of hexagonal boron nitride (hBN) and placed on top of a local, metal back gate. The structures were etched into a Hall geometry for initial transport measurements in Ref.~\onlinecite{Cao2018-bf}. However, in our capacitance measurements, we electrically short all contacts together to reduce the $RC$ charging time of the devices \footnote{See Supplemental Material for sample preparation details, which includes Refs.~\cite{Cao2016-ld,Cao2018-dj,Cao2018-bf,Wang614}}.

We apply an AC excitation to the MATBG contacts and a balancing AC excitation of variable phase and amplitude to a $\mathord{\sim}\SI{45}{\fF}$ reference capacitor connected to the back gate in a bridge configuration as shown in Fig.~\ref{schematic}. We measure small changes in the sample impedance by monitoring off-balance voltage accumulation at the balance point, and we model the total capacitance $C_\textrm{T}$ of the MATBG structure after subtraction of a constant background (see Supplemental Material for details) as consisting of two contributions: $C_\textrm{T}^{-1}= C_\textrm{geo}^{-1}+C_\textrm{q}^{-1}$. $C_\textrm{geo}$ is the geometric capacitance arising from the parallel plate geometry of the MATBG and local back gate while $C_\textrm{q}=Ae^2\partial n/\partial \mu$ is the quantum capacitance \cite{Luryi1988-su} which is directly proportional to the thermodynamic compressibility $\partial n/\partial \mu$ ($A$ is the lateral device area and $e$ is the elementary charge). By measuring modulation of the capacitance as a function of gate voltage and magnetic field, we detect the presence of gaps in the thermodynamic density of states. Importantly, we restrict our measurements to sufficiently low frequencies to ensure that modulation of the measured signal arises entirely from changes of the electronic compressibility and does not result from charging-rate effects \cite{Goodall1985-gn}. At high magnetic field, where the in-plane resistance becomes appreciable, we restrict ourselves to a qualitative discussion of the field-induced gaps \footnote{See Supplemental Material for measurement and analysis details, which includes Refs.~\cite{Eisenstein-1992,Yu2013-aa}}.

\begin{figure}[h!tp]
\centering
\includegraphics[scale=1]{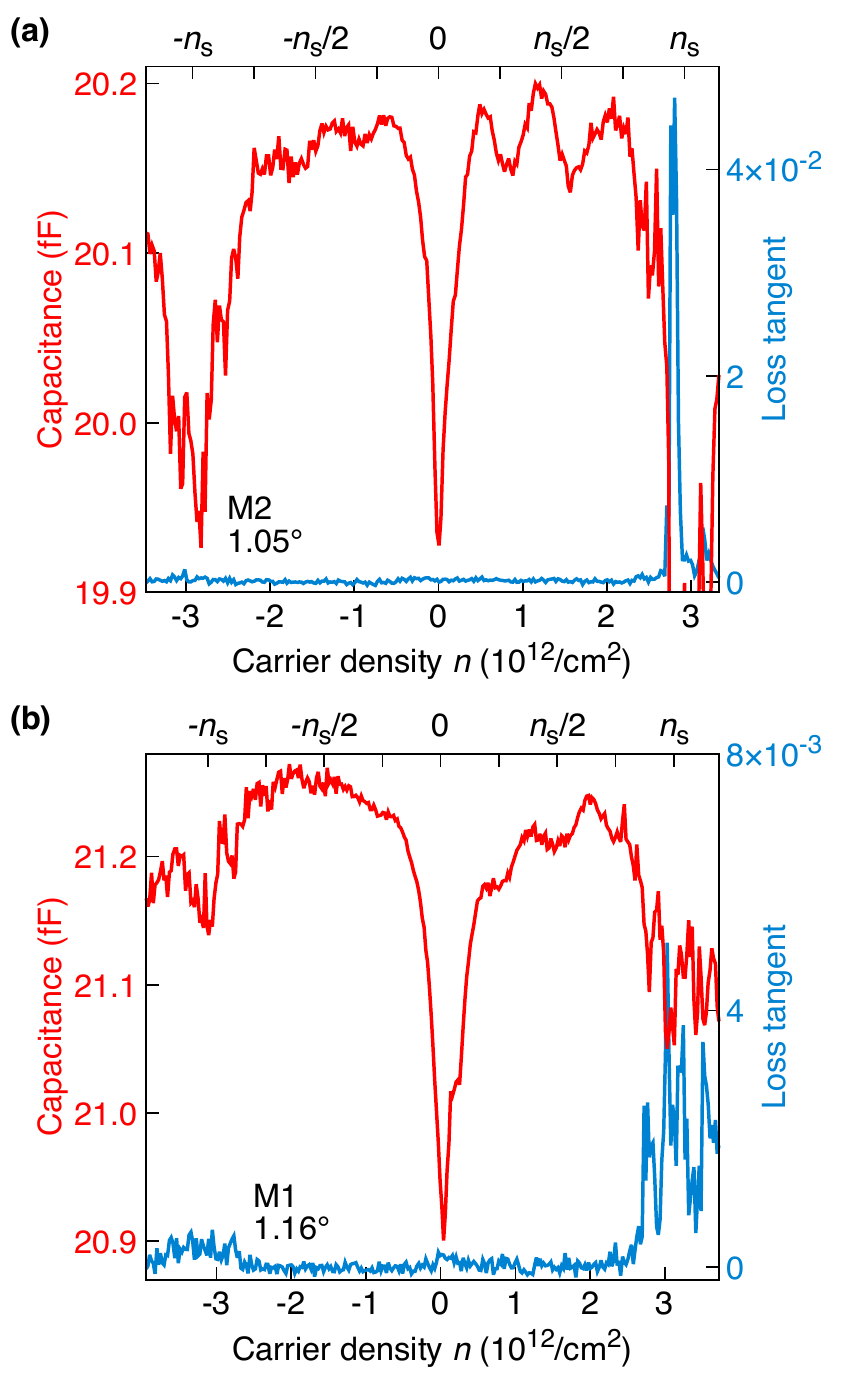}
\caption{\textbf{(a)} Plot of the capacitance (red trace) and loss tangent (blue trace) of device M2 as a function of carrier density at zero magnetic field and $\SI{280}{\milli\kelvin}$. Dips in the capacitance around $n_\textrm{s}/4$ and $n_\textrm{s}/2$ correspond to incompressible phases. Smaller dips around $-n_\textrm{s}/4$ and $-n_\textrm{s}/2$ also show incompressible features. \textbf{(b)} Capacitance and loss tangent of device M1 at zero magnetic field and $\SI{225}{\milli\kelvin}$. Similar incompressible features are seen at $n_\textrm{s}/4$ and $n_\textrm{s}/2$.}
\label{cap}
\end{figure}

Figure~\ref{cap}(a) and (b) show capacitance measurements (red traces) at zero magnetic field for the two devices (M2 and M1) that previously showed unexpected insulating and superconducting phases in Ref.~\onlinecite{Cao2018-bf}. Devices M2 and M1 were found previously to have twist angles of $\SI{1.05}{\degree}$ and $\SI{1.16}{\degree}$, respectively \cite{Cao2018-bf}. In both samples, we observe a Dirac-like feature at charge neutrality accompanied by broad, incompressible regions around $\SI{\pm3e12}{\per\centi\meter\squared}$, which correspond to either four electrons ($+n_\textrm{s}$) or four holes ($-n_\textrm{s}$) per moir\'{e} cell. We observe incompressible phases around $n_\textrm{s}/2$ (two electron per moir\'{e} cell) in both devices and a smaller feature around $-n_\textrm{s}/2$ in device M2. These incompressible features correspond to the previously reported insulating phases observed around $\pm n_\textrm{s}/2$ in Ref.~\onlinecite{Cao2018-bf}, however, in our measurement we find that the hole-doped state is significantly less incompressible despite comparable conductance values reported previously. Our results are consistent with more recent transport measurements in which resistive phases in the electron-doped regime are generally stronger than their hole-doped counterparts \cite{Yankowitz2018-qv, Sharpe-2019, Polshyn-2019, Efetov-2019}.

\begin{figure}[!htp]
\centering
\includegraphics[scale=1]{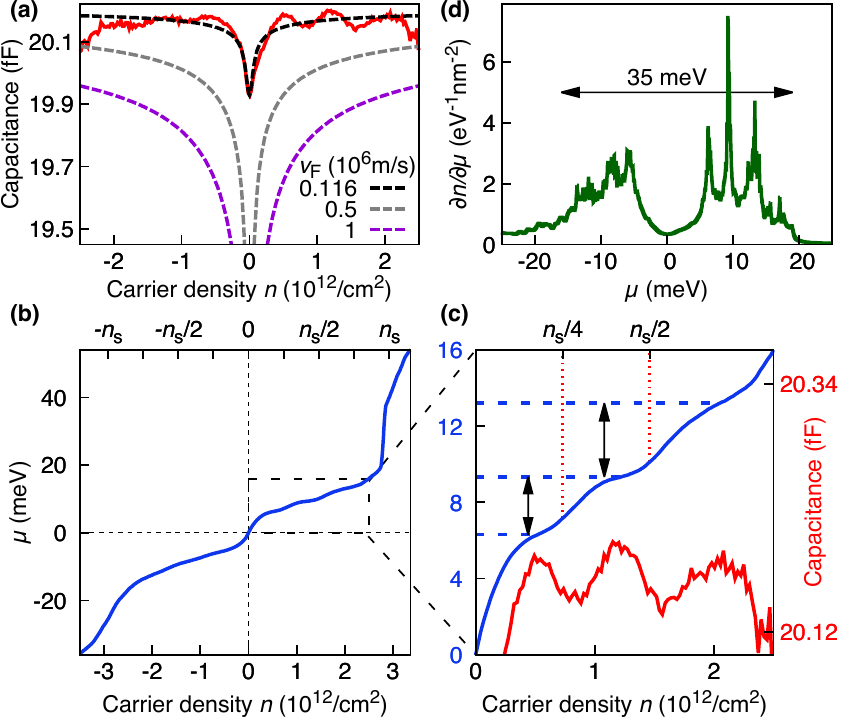}
\caption{\textbf{(a)} Fits of the capacitance data for sample M2. The Dirac point is best fit by $v_\textrm{F}=\SI{0.116e6}{\meter\per\second}$. \textbf{(b)} Shift of chemical potential of M2 as a function of carrier density calculated from integrating the inverse quantum capacitance. \textbf{(c)} Zoom-in to boxed region of (b). The increase in the chemical potential between the density of states maxima at both $n_\textrm{s}/4$ and $n_\textrm{s}/2$ is shown on the left axis with capacitance data from Fig.~\ref{cap}(a) replotted on the right axis. Arrows correspond to measured $\Delta \mu$ between density of states maxima. We find $\Delta_{n_\textrm{s}/4} = \SI{3.0\pm1.0}{\milli\electronvolt}$ and $\Delta_{n_\textrm{s}/2} = \SI{3.9\pm1.2}{\milli\electronvolt}$. \textbf{(d)} Extracted compressibility from the capacitance data in Fig.~\ref{cap}(a) as a function of the electronic chemical potential. Arrow is $\SI{35}{\milli\electronvolt}$ wide. }
\label{mu}
\end{figure} 

In both devices we find incompressible phases at $n_\textrm{s}/4$ and weak incompressible features at $-n_\textrm{s}/4$ in device M2. Importantly, the incompressible features at $n_\textrm{s}/4$ and $n_\textrm{s}/2$ are comparable in magnitude despite a lack of insulating temperature-dependence at $n_\textrm{s}/4$ reported previously \cite{Cao2018-bf,Yankowitz2018-qv,Polshyn-2019}. In device M2, raising the temperature to $\SI{5}{\kelvin}$, the highest temperature accessed, produced virtually no change in the capacitance features at commensurate filling (see Supplemental Material). We see no strong compressibility features around $\pm3n_\textrm{s}/4$ in contrast to recent transport studies \cite{Cao2018-bf,Yankowitz2018-qv,Sharpe-2019,Polshyn-2019,Efetov-2019}, although, we observe a gradual decrease in capacitance around $\pm3n_\textrm{s}/4$ as the system enters the superlattice band gaps which makes observation of incompressible features difficult to distinguish in this region. 

After subtracting a constant background capacitance $C_\textrm{back}$ (see Supplemental Material), the relationship
\begin{equation}
C_\textrm{T}^{-1} = C_\textrm{geo}^{-1}+\left(Ae^2\partial n/\partial \mu\right)^{-1}
\label{capeq}
\end{equation}
can be used to extract properties related to the compressibility $\partial n/\partial\mu$. At low temperatures the compressibility can be approximated well by the zero-temperature density of states. For a Dirac-like system with eight-fold degeneracy arising from spin, valley, and layer degrees of freedom, the density of states is given by 
\begin{equation}
\frac{\partial n}{\partial \mu}=\frac{4\left|E_\textrm{F}\right|}{\pi (\hbar v_\textrm{F})^2} = \frac{2\sqrt{2}}{\sqrt{\pi}\hbar v_\textrm{F}}\sqrt{|n|}.
\label{model}
\end{equation}
In order to take into account broadening due to the disorder profile across the lateral extent of the sample, we convolve Eq.~\ref{model} with a gaussian $g(n) = e^{-n^2/2\Gamma^2}/(\sqrt{2\pi}\Gamma)$ where $\Gamma$ characterizes the scale of the charge density broadening. We fit our capacitance data using Eq.~\ref{capeq} and $\partial n/\partial\mu * g(n)$ and determine $v_\textrm{F}$, $\Gamma$, and $C_\textrm{geo}$ from a best-fitting procedure. We estimate the area $A$ from the lithographic dimensions of the MATBG flake that lie over the back gate. The capacitance data from sample M2 is best fit by $v_\textrm{F}=\SI{0.116e6}{\meter\per\second}$ and $\Gamma=\SI{4.0e10}{\per\centi\meter\squared}$, both of similar magnitude to the values extracted previously with a different sample ($\SI{1.08}{\degree}$) \cite{Cao2018-dj}. Figure~\ref{mu}(a) shows the capacitance data from M2 overlaid with the model evaluated at various values of $v_\textrm{F}$.  

The shift of the chemical potential as carriers are added to the device can be extracted by integrating the inverse quantum capacitance $\Delta \mu = \int \frac{\partial \mu}{\partial n}dn = \int Ae^2 \left(C_\textrm{T}^{-1} - C_\textrm{geo}^{-1}\right)dn$. Figure~\ref{mu}(b) shows the calculated chemical potential as a function of carrier density for device M2. Portions of the trace with relatively flat slope correspond to compressible phases where the chemical potential shifts relatively little as carriers are added to the system whereas steeper regions correspond to reductions in the density of states. The superlattice band gaps around $\pm n_\textrm{s}$ manifest as the steep slopes near $n=\SI{\pm3e12}{\per\centi\meter\squared}$.  On the electron side, where the size of the capacitance dips is appreciable, the thermodynamic gap between the density of states peaks can be calculated for both $n_\textrm{s}/4$ and $n_\textrm{s}/2$ as shown in Fig.~\ref{mu}(c). We find $\Delta_{n_\textrm{s}/4}=\SI{3.0\pm 1.0}{\milli\electronvolt}$ and $\Delta_{n_\textrm{s}/2}=\SI{3.9\pm 1.2}{\milli\electronvolt}$, where the error associated with the gap estimation arises from a systematic error in the determination of $C_\textrm{geo}$. See Supplemental Material for a discussion of the error estimation and its propagation in the thermodynamic gap measurements. The gap at $n_\textrm{s}/4$ was either not observed previously \cite{Cao2018-dj,Cao2018-bf,Kerelsky2018-py,Choi-2019}, found to have non-activated temperature dependence  \cite{Yankowitz2018-qv,Polshyn-2019}, or the resistive feature was not discussed in detail \cite{Sharpe-2019}. A recent transport study reported simply-activated temperature dependence at quarter-filling with a gap value of $\SI{0.14}{\milli\electronvolt}$, though the presence of a gap at charge neutrality may indicate that the twisted bilayer graphene is aligned with the hBN substrate \cite{Efetov-2019}. Our estimate of $\Delta_{n_\textrm{s}/2}$ is significantly larger than the previously reported values of $\SI{0.31}{\milli\electronvolt}$ in Ref.~\onlinecite{Cao2018-dj}, $\mathord{\sim}\SI{1.5}{\milli\electronvolt}$ in Ref.~\onlinecite{Yankowitz2018-qv}, and $\SI{0.37}{\milli\electronvolt}$ in Ref.~\onlinecite{Efetov-2019}

We expect the gap extracted from thermodynamic compressibility to be larger than the activation gap measured through the temperature dependence of the resistivity. As temperature is increased, the electron--electron correlations that create the many-body gap may weaken, causing the gap to decrease as a function of temperature and leading to an underestimation in activation measurements. By measuring at a fixed, low temperature, the gap derived from compressibility is potentially larger. Additionally, there may be a large density of charge carriers that can be thermally excited across the many-body gap at energies that are closer to the Fermi level than the density of states maxima, leading to a smaller activation gap. If we measure the shift in chemical potential just around the steepest portions of trace in Fig.~\ref{mu}(c), we find values of approximately $\SI{2}{\milli\electronvolt}$ and $\SI{2.5}{\milli\electronvolt}$ for the $n_\textrm{s}/4$ and $n_\textrm{s}/2$ states, respectively, which may compare more directly to activation measurements.

We can also make a comparison to recent STS measurements in which splittings of the van Hove singularity at $n_\textrm{s}/2$ of roughly $\SI{7.5}{\milli\electronvolt}$ \cite{Kerelsky2018-py} and $4-\SI{8}{\milli\electronvolt}$ \cite{Choi-2019} have been measured. Because the STM tip is placed over a clean, atomically resolved region of the sample, the effects of disorder averaging are avoided, leading to a potentially larger observed spectroscopic gap. We note that the van Hove singularity separation when the Fermi level lies at half-filling as seen in STS measurements differs qualitatively from the chemical potential separation observed in compressibility. In the latter case, the carrier density and the band structure itself vary as the Fermi level is raised due to density-dependent electron--electron interactions. Additionally, STS measures the single particle density of states which is a different quantity from the thermodynamic density of states $\partial n/\partial \mu$ accessed in our measurements.

We plot the compressibility $\partial n/\partial\mu =\frac{1}{Ae^2}\left(C_\textrm{T}^{-1}- C_\textrm{geo}^{-1}\right)^{-1}$ as shown in Fig.~\ref{mu}(d). The vertical scale of the compressibility is very sensitive to the precise value of $C_\textrm{geo}$, particularly the highly compressible phases in which $C_\textrm{T}\approx C_\textrm{geo}$. Nonetheless, the horizontal axis is much less sensitive to variation in $C_\textrm{geo}$, and we estimate the bandwidth of the two low-energy moir\'{e} bands as $\SI{35}{\milli\electronvolt}$. If we vary the precise value of $C_\textrm{geo}$ over an estimated uncertainty (see Supplemental Material for details), the bandwidth varies from as small as $\SI{25}{\milli\electronvolt}$ to as large as $\SI{45}{\milli\electronvolt}$. This range of values is much larger than initial calculations for a rotation angle of $\SI{1.05}{\degree}$ \cite{Morell-2010,Bistritzer2011-fm,Cao2018-dj}, but it is consistent with the $\SI{41}{\milli\electronvolt}$ separation of the valence and conduction band van Hove singularities predicted by recent tight-binding calculations and the $\SI{55}{\milli\electronvolt}$ separation observed by STS for a slightly larger rotation angle of $\SI{1.10}{\degree}$ \cite{Kerelsky2018-py}. 

 \begin{figure}[h!tp]
\centering
\includegraphics[scale=1]{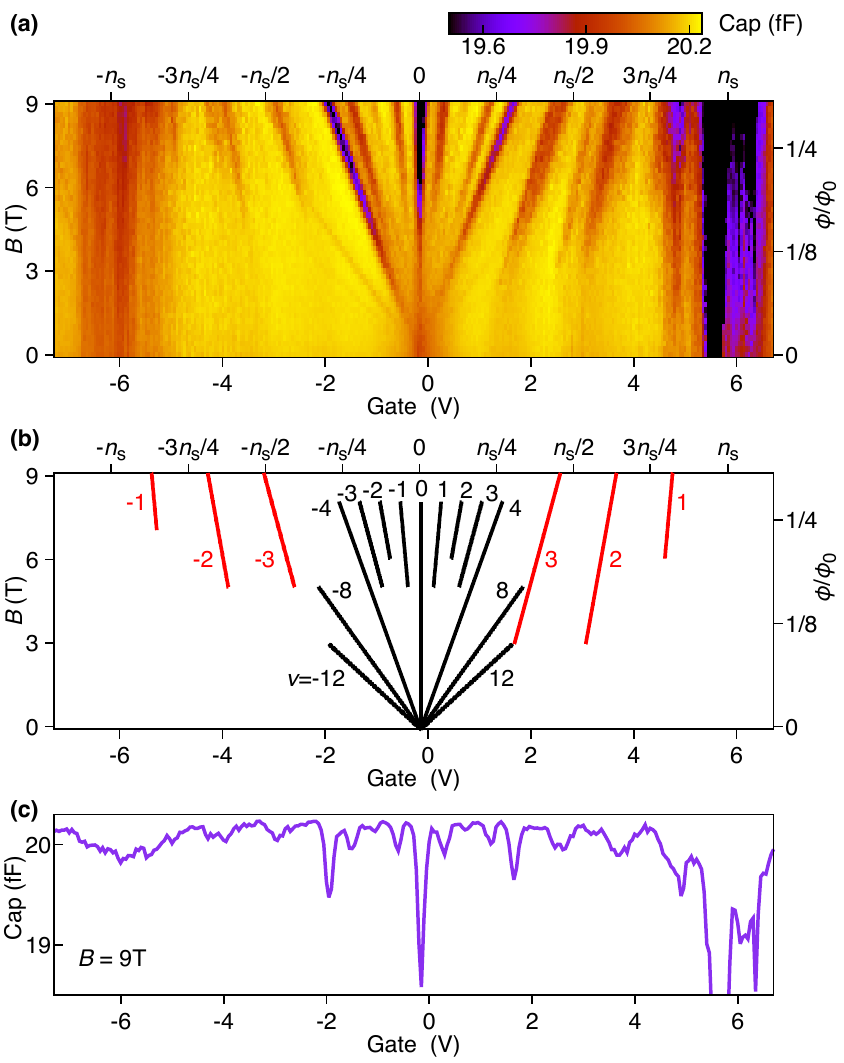}
\caption{\textbf{(a)} Capacitance as a function of gate voltage and perpendicular magnetic field. Color scale has been suppressed below $\SI{19.5}{\fF}$ in order to show more detail. \textbf{(b)} Map of observed gaps in magnetic field measurement in (a). Black traces indicate cyclotron and exchange gaps arising from charge neutrality with the filling factor labeled. Red traces indicate gaps emanating from high magnetic field due to ``Hofstadter'' replica minibands. \textbf{(c)} Capacitance linecut at $B=\SI{9}{\tesla}$ from panel (a).}
\label{field}
\end{figure}

We also measure the evolution of the compressibility with magnetic field up to the quantum Hall regime. In Fig.~\ref{field}(a) we plot the capacitance as a function of gate voltage and magnetic field. The incompressible phases at commensurate filling of the miniband do not appear to change with perpendicular magnetic field up to about $\SI{3}{\tesla}$ after which field-induced gaps arise and coexist, making it difficult to track the relatively broader features associated with the zero-field incompressible phases. The most prominent feature in Fig.~\ref{field}(a) is a four-fold degenerate Landau fan that emerges from charge neutrality. Despite anticipating an eight-fold degenerate zero energy LL arising from spin, valley, and layer degrees of freedom, our system never develops a compressible phase at charge neutrality, indicating that layer or valley symmetry breaking is present even at zero magnetic field. Four-fold degeneracy may be evidence of $C_3$ rotational symmetry breaking as recently proposed \cite{Zhang-2019-ll}. At larger magnetic fields we observe incompressible phases at filling factors within the lowest LL octet. These additional incompressible phases presumably arise from exchange-driven gaps as reported in monolayer and bilayer graphene \cite{Zhang2006-xh,Jiang2007-ze,Checkelsky2008-is,Zhao2010-sq,Young2012-yh,Hunt2017-mu}. In addition, LLs emerging from the superlattice gaps of device M2 are apparent in Fig.~\ref{field}(a) as well as from device M1 in Fig.~S6(a) in the Supplemental Material. 
 
Additionally, we see a set of gaps emerging from high magnetic field whose intercepts terminate near the fractional filling densities as indicated by the red traces in Fig.~\ref{field}(b). These gaps appear to form as a result of ``Hofstadter'' fractal minibands \cite{Hofstadter1976-yj,Wannier1978-dr}. Similar features have been observed extensively in aligned graphene--hBN structures \cite{Dean2013,Ponomarkenko-2013,Hunt1237240} as well as twisted bilayer graphene devices \cite{Cao2016-ld,Kim2017-os}. In Fig.~\ref{field}(a) some of the gaps which approach commensurate filling appear doubled, suggesting there may be multiple regions within the device with slightly different twist angle. The number of flux quanta per moir\'{e} unit cell is plotted on the right vertical axes in Fig.~\ref{field}(a) and (b). The incompressible phases shown in red in panel Fig.~\ref{field}(b) intersect at $\phi/\phi_0 = 1$ which occurs at $B=\SI{29.64}{\tesla}$ for $\theta=\SI{1.05}{\degree}$. Unlike Ref.~\onlinecite{Yankowitz2018-qv}, where no Landau fan was observed emanating from $-n_\textrm{s}/4$, we observe an incompressible phase emerge from high magnetic field whose intercept terminates at $-n_\textrm{s}/4$. Additionally, we observe an incompressible phase emanating from $n_\textrm{s}/4$ corresponding to $\nu=3$ and not $\nu=1$ as seen previously \cite{Yankowitz2018-qv}. See Supplemental Material for a Wannier diagram and loss tangent in Fig.~S5 as well as the field dependence of device M1 with similar features in Fig.~S6. The presence of strong quantum oscillations at low magnetic field close to commensurate filling in transport \cite{Cao2018-bf,Yankowitz2018-qv,Efetov-2019} indicates the formation of an emergent Fermi surface and its quantization in magnetic field. Oscillations in transport measurements reflect the field dependence of both the scattering time $\tau$ as well as the density of available charge carriers. Because low-frequency capacitance measurements are insensitive to changes in the scattering rate, strong features are not expected when Landau quantization is weak which may explain the lack of low-field capacitance oscillations in our data in contrast with transport \cite{Cao2018-bf,Yankowitz2018-qv}. 

Although there is no consensus on the nature of the commensurate insulating phases, our results allow us to comment on a few recent proposals. One effort posits that the correlated insulating phase at half-filling is not a Mott-like insulator, but rather a Wigner crystal in which the electrons freeze into an emergent lattice as a result of long-range Coulomb repulsion \cite{Padhi2018-un, Padhi-2018-2}. Although a Wigner crystal is electrically insulating due to pinning of the electron lattice by disorder, domains, or the moir\'{e} lattice itself, the compressibility of the Wigner crystal is expected to be large and negative due to long-range Coulomb interactions \cite{Bello-1981}. Unlike DC transport, compressibility is one of the few techniques which is capable of providing positive evidence of Wigner crystallization. In our measurements, the compressibility decreases at commensurate filling while remaining positive and non-diverging, implying the likely formation of an energy gap and not a highly (negatively) compressible phase expected for an ideal Wigner crystal. However, unlike the conventional case, if the moir\'{e} potential strongly pins the electron lattice, it may be possible to form a thermodynamic energy gap \cite{Padhi-convo}. We also cannot rule out the possibility of such a Wigner crystal and another gapped phase coexisting via phase separation as has been speculated to occur in GaAs bilayers \cite{Zhang-2014-wc}. Additionally, it may be possible to interpret the multiple density of states peaks as arising from differential strain between the two twisted bilayer graphene layers \cite{Bi-2019-strain}.

Although many band structure calculations predict narrow low energy bands between $5$ and $\SI{10}{\milli\electronvolt}$ \cite{Morell-2010,Bistritzer2011-fm,Trambly2010-ui,Cao2018-dj,Vafek-2019}, recent focus on lattice relaxation effects have brought estimates closer to experiment ($\mathord{\sim}\SI{20}{\milli\electronvolt}$) \cite{Carr-2019-lattice,Walet-2019-lattice}, however, at least one other lattice relaxation model predicts a narrowing of the bandwidth as compared to unrelaxed calculations \cite{Lucignano-2019-lattice}. Our data support a bandwidth in the range of $25$ to $\SI{45}{\milli\electronvolt}$, suggesting that the non-interacting band structure is not as narrow as anticipated by theory, leading to larger values of kinetic energy. This suggests that, in the creation of the correlated insulating states, the kinetic energy may play a more substantive role than many single-particle calculations imply. Moreover, as recently proposed \cite{Yuan-2019-vh}, it is possible that the interaction effects are also strongly enhanced by a power-law diverging van Hove singularity.

In summary, we used compressibility measurements to access the shift of the chemical potential as the low-energy band structure is filled. We report a reduced Fermi velocity, a wide $\mathord{\sim}\SI{35}{\milli\electronvolt}$ bandwidth compared with many electronic structure calculations, and measure the gap widths at $n_\textrm{s}/4$ and $n_\textrm{s}/2$. The incompressible features at commensurate filling show essentially no field evolution up to $\SI{3}{\tesla}$ before becoming obscured by other field-induced gaps. We do not observe strong Landau quantization at low magnetic field around commensurate filling, but at larger magnetic field we detect ``Hofstadter'' gaps that differ from previous transport studies \cite{Cao2018-bf,Yankowitz2018-qv,Efetov-2019}.

\begin{acknowledgments}
We acknowledge useful discussions with Brian Skinner, Stevan Nadj-Perge, Philip W. Phillips, Bikash Padhi, Liang Fu, A.F. Young, V. Fatemi, D.S. Wei, and J.D. Sanchez-Yamagishi. In this work, sample fabrication (Y.C. and P.J.H.) was primarily supported by the National Science Foundation (NSF: DMR-1809802) and the Gordon and Betty Moore Foundation's EPiQS Initiative through grant GBMF4541. Capacitance measurements (S.L.T., A.D and R.C.A.) were supported by the STC Center for Integrated Quantum Materials, NSF grant number DMR-1231319. K.W. and T.T. acknowledge support from the Elemental Strategy Initiative conducted by the MEXT, Japan, A3 Foresight by JSPS and the CREST (JPMJCR15F3), JST. This work made use of the Materials Research Science and Engineering Center Shared Experimental Facilities supported by the NSF (DMR-0819762) and of Harvard's Center for Nanoscale Systems, supported by the NSF (ECS-0335765).
\end{acknowledgments}

\bibliography{magicAngle}

\newcommand{\beginsupplement}{%
		\setcounter{table}{0}
		\renewcommand{\thetable}{S\arabic{table}}%
		\setcounter{figure}{0}
		\renewcommand{\thefigure}{S\arabic{figure}}%
		\setcounter{equation}{0}
		\renewcommand{\theequation}{S\arabic{equation}}
	}
		\beginsupplement
		
\section{Supplemental Material}

\subsection{Sample preparation}
The samples were fabricated using a dry transfer technique modified with a rotation stage described previously \cite{Cao2016-ld,Cao2018-dj,Cao2018-bf} and originally measured in Ref.~\onlinecite{Cao2018-bf}. Monolayer graphene and hexagonal boron nitride (hBN) crystals of $10$--$\SI{30}{\nano\meter}$ thickness were exfoliated onto clean $\ce{SiO2}/\ce{Si}$ substrates, identified optically, and characterized with atomic force microscopy. The twisted bilayer graphene was constructed by ``tearing and stacking'' a clean monolayer with a precise rotational misalignment. A poly(bisphenol A carbonate) (PC)/polydimethylsiloxane (PDMS) stack on a glass slide was used to first pick up a piece of hBN at $\SI{90}{\celsius}$ after which the van der Waals forces between hBN and graphene were used to tear a graphene flake close to room temperature. The separated graphene pieces were rotated manually by a twist angle around $1.2$--$\SI{1.3}{\degree}$ and stacked on top of one another. The stack was encapsulated by a piece of hBN on the bottom and released  at $\SI{160}{\celsius}$ onto a Cr/PdAu metal back gate on top of a highly resistive $\ce{SiO2}/\ce{Si}$ substrate. The samples were not annealed to prevent the twisted bilayer graphene from relaxing back to Bernal-stacked bilayer graphene. The device geometry was defined using standard electron-beam lithography techniques and reaction ion etching with fluoroform and $\ce{O2}$ plasmas. Electrical contact was made to the MATBG with Cr/Au edge-contacted leads \cite{Wang614}. 

\subsection{Measurements}

\begin{figure*}[h!tp]
\centering
\includegraphics[scale=1]{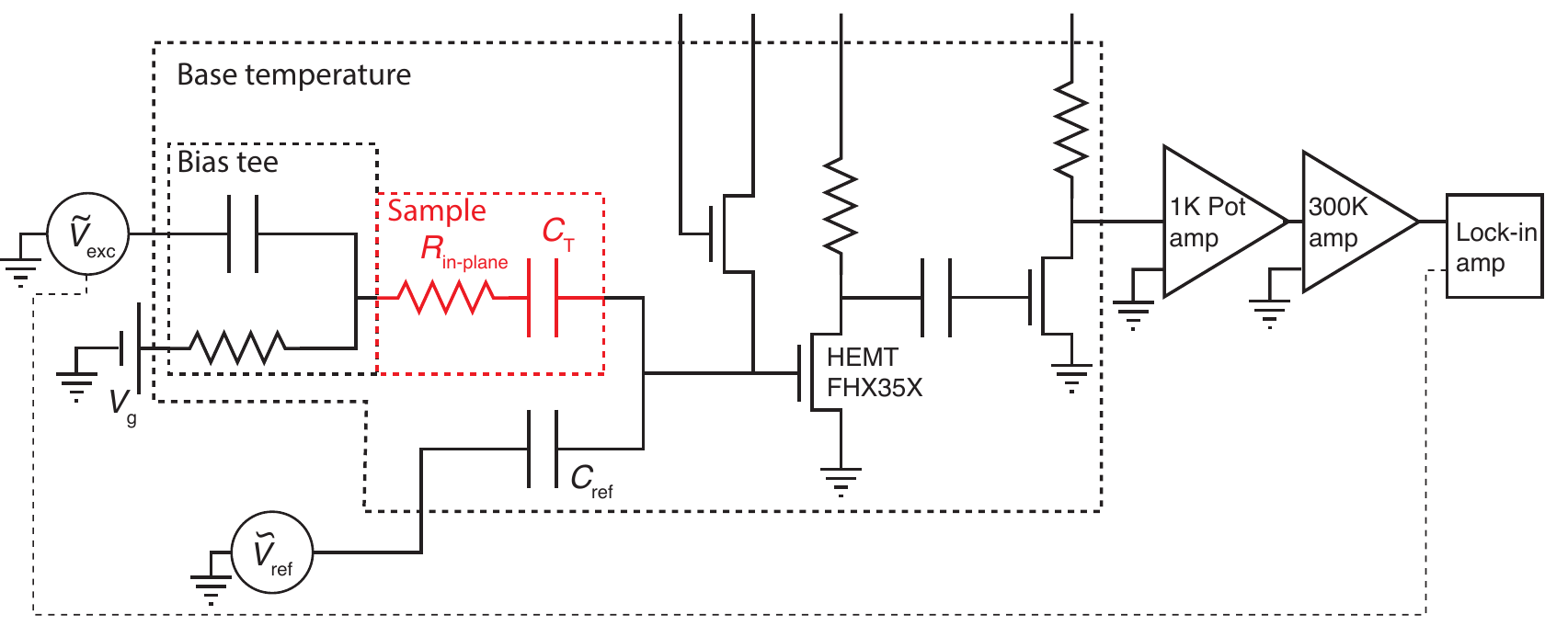}
\caption{Circuit for the capacitance and loss tangent measurements.}
\label{circuit}
\end{figure*}

Sample M2 was measured in a helium-3 cryostat at $\SI{280}{\milli\kelvin}$, and sample M1 was measured in a dilution refrigerator at a temperature of $\SI{225}{\milli\kelvin}$ with the exception of the high temperature measurements of M2. Capacitance and loss tangent measurements were carried out on homemade cryogenic capacitance bridges on the same chip carriers as the samples. Figure~\ref{circuit} shows the full circuit schematic. The sample is modeled as a series resistor and capacitor in red. AC and DC voltages are applied to the sample through a cryogenic bias tee adjacent to the sample. An AC signal of variable phase and amplitude is applied to a fixed $\mathord{\sim}\SI{45}{\fF}$ reference capacitor to balance the sample capacitance at the bridge balance point. After an initial balance is achieved, changes in the sample capacitance are inferred from the off-balance voltage that accumulates at the balance point. The size of the reference capacitor is determined on a subsequent cooldown with a known $\SI{2}{\pF}$ capacitor. Three Fujitsu FHX35X high electron mobility transistors (HEMTs) are used in a double-amplifier configuration. The main amplification stage occurs at the gate of the labeled HEMT which has been cleaved in half to minimize its stray capacitance. A second HEMT (also cleaved) is used as a variable resistor (typically set to about $\SI{100}{\mega\ohm}$) to pinch off the measurement HEMT's channel to about $\SI{100}{\kilo\ohm}$. A third follower HEMT which is uncleaved drives the signal to a homemade wide-bandwidth amplifier located at the $\SI{1}{\kelvin}$ pot of the fridge followed by a similar amplification stage at room temperature before being measured on a lock-in amplifier (Signal Recovery SR7280). An excitation voltage of $\SI{2.8}{\milli\volt}$ RMS was applied at $\SI{150}{\kilo\hertz}$ for all measurements unless otherwise indicated.

For a sample that can be modeled as a series resistor and capacitor, the output of the capacitance bridge has two components: an in-phase and out-of-phase component. The in-phase component $X$ is given by 
\begin{equation}
	X = \frac{-C_\textrm{T}/C_\textrm{ref}}{1+\left(\omega R_\textrm{in-plane} C_\textrm{T}\right)^2}
\end{equation}
while the out-of-phase component $Y$ is given by
\begin{equation}
	Y = \frac{\omega R_\textrm{in-plane} C^2_\textrm{T}/C_\textrm{ref}}{1 + \left(\omega R_\textrm{in-plane}C_\textrm{T}\right)^2}
\end{equation}
where $C_\textrm{ref}$ is the value of the fixed reference capacitor, $R_\textrm{in-plane}$ represents the effective series resistance of the MATBG, $C_\textrm{T}$ the total sample capacitance (which includes both the geometric and quantum contributions), and $\omega$ is related to the measurement frequency $f = \omega/2\pi$. The loss tangent is defined as the ratio between the resistive and reactive impedances of the sample $\frac{Z_\textrm{res}}{Z_\textrm{react}} = \frac{R_\textrm{in-plane}}{1/(\omega C_\textrm{T})}$. This quantity is exactly equal to the ratio of $Y$ to $-X$: 
\begin{equation}
	\textrm{loss tangent} = \frac{Y}{-X} = \omega R_\textrm{in-plane}C_\textrm{T}.
\end{equation}
In the thermodynamic limit, $\omega$ satisfies $\omega \ll \frac{1}{R_\textrm{in-plane}C_\textrm{T}}$. In other words, the excitation frequency is slow enough that the sample has time to completely charge on each cycle of the excitation. This allows a decoupling of the capacitive and resistive impedances in the $X$ channel: 
\begin{equation}
	X \approx \frac{-C_\textrm{T}}{C_\textrm{ref}}
\end{equation}
and allows us to compute the total capacitance $C_\textrm{T}$ in units of the reference.

In our analysis the role of the loss tangent is to confirm that our measurements are in the thermodynamic (low frequency) limit. Whenever we see modulation of the in-phase component accompanied by a featureless loss tangent, we know that $\omega \ll \frac{1}{R_\textrm{in-plane}C_\textrm{T}}$ and modulation of the in-phase channel can be interpreted as modulation of the sample's capacitance (and hence changes in the compressibility). In Fig.~2(a) and (b) the loss tangent remains near zero and perfectly flat between the superlattice gaps at $\pm n_\textrm{s}$. This confirms that the changes in the capacitance at commensurate filling arise from the compressibility and not a failure of the samples to charge completely. This rules out localization causing a large in-plane resistance without an attendant decrease in the thermodynamic density of states. The loss tangent spikes suddenly as the superlattice gaps are entered indicating a sudden increase in the in-plane resistance. This is accompanied by a sharp decrease in the capacitance. The decrease in capacitance is in part due to a decrease in the compressibility as the Fermi level enters a bandgap, but it is also related to a failure of the sample to charge completely at the given measurement frequency. In this limit the term $\left(\omega R_\textrm{in-plane} C_\textrm{T}\right)^2$ dominates the denominator of $X$ and limits our ability to be quantitative about the compressibility. 

The same is true at high magnetic field when the Fermi level lies in a cyclotron or exchange gap. The substantial reduction in the density of states causes the in-plane resistance to dominate our measurement. This does not affect a qualitative discussion of the compressibility evolution in magnetic field (e.g. the emergence of exchange gaps or trajectories of insulating features), but does preclude a quantitative measure of the compressibility.

\subsection{Capacitance corrections and background subtraction}
Due to slight mismatches in the cabling of the reference and sample lines due to different attenuators as well as slight offsets in the relative phase of the excitation and reference voltages sources, our measurements show a constant offset in the out-of-phase component (which for small values of $Y$ is approximately equal to the loss tangent) that is not attributed to the sample. In order to correct for this small phase shift we find the values of the in-phase and out-of-phase components of our signal near a highly compressible state that should have minimal in-plane resistance and an extremely small out-of-phase component. We apply a rotation matrix $M(\theta)$ (typically the rotation angle $\theta \approx \SI{1}{\degree}$) to our in- and out-of-phase signal measurements to correct this artifact. This leaves the in-phase component virtually unchanged and shifts the out-of-phase signal to the correct baseline. 

Our samples have a substantial stray capacitance of around $\SI{50}{\fF}$ that manifests in our measurement as a constant background  in parallel with the sample $C_\textrm{raw} = C_\textrm{back} + C_\textrm{T} = C_\textrm{back} + \left(C_\textrm{geo}^{-1}+C_\textrm{q}^{-1}\right)^{-1}$. In order to accurately subtract this constant background, we utilize the field-dependent capacitance measurements where we can accurately determine the minima of the cyclotron gaps of the Landau fan emerging from charge neutrality (see Fig.~4(a)). Regardless of the underlying band structure, all Landau levels can be characterized by a field-dependent orbital degeneracy $\phi/\phi_0 = BAe/h$ where $\phi$ is the total magnetic flux through the sample, $\phi_0$ is the flux quantum, $B$ is the magnetic field, $A$ the sample area, $e$ is the elementary charge, and $h$ is Planck's constant. This orbital degeneracy is augmented by a factor of $8$ arising from the spin, valley, and layer degrees of freedom. Therefore, between the filling factors $\nu = \pm 4$, we know the total charge accumulated in the sample is given by $8BAe^2/h$. The total charge accumulated in the sample is also given by integrating the total capacitance: $Q = \int_{\Delta V} C_\textrm{T} dV$ where the limits of integration are determined by the gate voltages in Fig.~4(a). Therefore, the appropriate value of $C_\textrm{back}$ is found by enforcing 
\begin{equation}
	8BAe^2/h = Q = \int_{\Delta V} \left(C_\textrm{raw} - C_\textrm{back}\right)dV.
\end{equation}

In this analysis we assume that the strongest gaps emerging from charge neutrality correspond to $\nu=\pm4$ as expected for a twisted bilayer graphene system. This is confirmed by calculating the slope of the gaps in Fig.~4(a) and using the relationship
\begin{equation}
	\nu = \frac{nA}{\phi/\phi_0} = \frac{n\phi_0}{B} = \frac{\left(\overline{C_\textrm{T}}/A\right)\Delta V\phi_0}{eB}
\end{equation}
where $\overline{C_\textrm{T}}$ is the average total capacitance between the LL minima. Because $\overline{C_\textrm{T}}$ is roughly equal to the geometric value of the capacitance which is well approximated by a parallel plate model, we can say $\overline{C_\textrm{T}}/A \approx \epsilon \epsilon_0/d$ where $\epsilon$ is the relative dielectric of the hBN ($\mathord{\sim} 4.5$), $d$ is the thickness of the dielectric ($\mathord{\sim}\SI{30}{\nano\meter}$) determined from atomic force microscopy, and $\epsilon_0$ is the vacuum permittivity, yielding: 
\begin{equation}
\nu \approx \frac{\epsilon \epsilon_0\Delta V\phi_0}{deB}.
\end{equation}
After extracting the slope of the gap in Fig.~4(a) and equating it to $\frac{B}{\Delta V}$ and using estimated values for $\epsilon$ and $d$, we can verify that $\nu = \pm4$. Additionally this allows us to verify that our background subtraction is reasonable by confirming $\overline{C_\textrm{T}} \approx C_\textrm{raw} - C_\textrm{back}$.

 \subsection{Converting from gate voltage to carrier density}
Unlike transport measurements, our capacitance technique allows us to convert gate voltage to carrier density exactly. Typically, in transport the gating capacitance is taken as a constant $\overline{C_\textrm{T}}$ (typically extracted from Landau fans or modeled with parallel plate geometries) and is often described as purely geometric but in reality is an average value of the {\em total} capacitance that includes contributions from the quantum capacitance that vary as a function of density. In most samples $C_\textrm{q} \gg C_\textrm{geo}$ so that $C_\textrm{T} \approx C_\textrm{geo}$, allowing this approximation to hold. For our measurements we can simply integrate the total capacitance with respect to gate voltage to directly calculate the induced charge density: 
\begin{equation}
	n(V) = \frac{1}{Ae}\int_{V_\textrm{Dirac}}^V C_\textrm{T}(V') dV'
\end{equation}
where we have set the carrier density at the gate voltage associated with the Dirac point to $0$. For our samples the quantum capacitance $C_\textrm{q}$ is always much larger than the geometric capacitance  inside the superlattice gaps. Therefore, the relationship between carrier density and voltage is roughly proportional, but there are subtle nonlinearities near locations of relatively small quantum capacitance (e.g.~near charge neutrality) that are captured in this conversion.

 \subsection{Determining the geometric capacitance}
\begin{figure}[h!tp]
\centering
\includegraphics[scale=1]{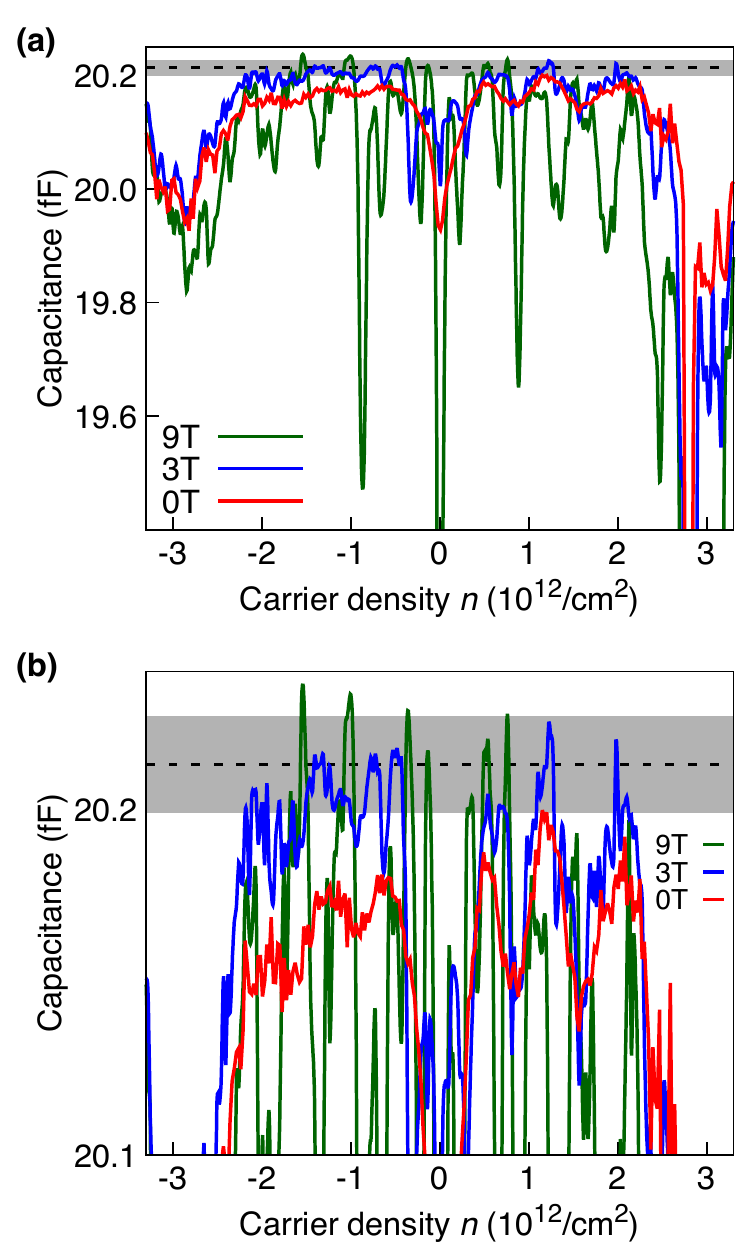}
\caption{\textbf{(a)} Plot of capacitance traces at $B=0$ (red), $\SI{3}{\tesla}$ (blue), and $\SI{9}{\tesla}$ (green) as well as the estimate for $C_\textrm{geo}$ (black dashed trace). The gray region represents the estimated uncertainty in $C_\textrm{geo}$. \textbf{(b)} Zoom-in of (a).}
\label{geo}
\end{figure}

Our quantitative analysis relies on estimating the value of the geometric capacitance $C_\textrm{geo}$. In order to estimate $C_\textrm{geo}$ we use the model $\mathcal{C}(n)$ for the total capacitance:
\begin{equation}
	\mathcal{C}(n) = \left(\frac{1}{C_\textrm{geo}} + \frac{1}{Ae^2\partial n/\partial\mu(n)}\right)^{-1}.
\end{equation}
For a bilayer graphene system with eight-fold degeneracy the density of states is given by
\begin{equation}
	\frac{\partial n}{\partial \mu} = \frac{4\left|E_\textrm{F}\right|}{\pi\left(\hbar v_\textrm{F}\right)^2} = \frac{2\sqrt{2}}{\sqrt{\pi}\hbar v_\textrm{F}}\sqrt{|n|}
\end{equation}
where $E_\textrm{F}$ is the Fermi energy, $v_\textrm{F}$ the Fermi velocity, $\hbar$ is Planck's constant. Additionally, we take into account the spatial broadening due to the disorder profile by convolving $\partial n/\partial \mu$ with a gaussian $g(n) = e^{-n^2/2\Gamma^2}/(\sqrt{2\pi}\Gamma)$ where $\Gamma$ characterizes the scale of the disorder broadening. We fit $\partial n/\partial \mu * g(n)$ to our data to determine best-fit values of $v_\textrm{F}$, $\Gamma$, and $C_\textrm{geo}$. The value of $C_\textrm{geo}$ from best-fitting agrees nicely with the peaks in the highly compressible Landau levels that we expect to be very close to the geometric capacitance and possibly in excess if negative compressibility is present \cite{Eisenstein-1992,Yu2013-aa}. See Fig.~\ref{geo}(a) and (b) for plots of $C_\textrm{geo}$ overlaid with the field-dependent capacitance data. In order to account for the fact that our fit-derived value may deviate from the true value of the geometric capacitance, we estimate an uncertainty $\delta c = \SI{0.014}{\fF}$ in $C_\textrm{geo}$ based on a visual analysis of the compressible Landau level peaks. The lower bound of our uncertainty corresponds to assuming that the density of states maxima in the zero field capacitance data are nearly perfectly compressible. This is a reasonable lower bound assuming that the density of states peaks do not exhibit negative compressibility. This is justifiable if we compare these maxima to the highly compressible Landau levels between $\nu=-12$ and $\nu=-4$ at $B=\SI{3}{\tesla}$, a large enough field for good Landau quantization but low enough that the exchange gaps and ``Hofstadter'' features at high magnetic field do not overlap. The capacitance signal forms clear plateaus with no sign of negative compressibility and remains larger in value than the zero-field data at all densities. See panel (b) of Fig.~\ref{geo} where the blue trace saturates close to the fit-derived value of $C_\textrm{geo}$ between about $n = -1$ and $\SI{-0.5e12}{\per\centi\meter\squared}$ and remains larger than all capacitance values in the red trace. The upper bound of $C_\textrm{geo}$ is placed near the highest capacitance values recorded at high magnetic field where we expect the capacitance peaks to be highly compressible and possibly enhanced beyond the geometric value if negative compressibility is present. The role of the geometric capacitance uncertainty and its propagation in the thermodynamic gap and bandwidth calculations are detailed below.

\subsection{Uncertainty in thermodynamic gaps}
The inverse compressibility is integrated to extract the chemical potential $\mu$ as a function of carrier density $n$. A small error (compared to the magnitude of $C_\textrm{geo}$) in the geometric capacitance causes a spurious linear background in the overall slope of $\mu(n)$. If we take the true geometric capacitance to be $C_\textrm{geo}$ and $\delta c$ a small error we compute
\begin{equation}
	\frac{1}{C_\textrm{T}} - \frac{1}{C_\textrm{geo} + \delta c} \approx \frac{1}{C_\textrm{T}} - \frac{1}{C_\textrm{geo}} + \frac{\delta c}{C_\textrm{geo}^2}.
	\label{error1}
\end{equation}
Multiplying through by $Ae^2$, we can cast Eq.~\ref{error1} in terms of the inverse compressibility and an associated deviation:
 \begin{equation}
 	Ae^2\left(\frac{1}{C_\textrm{T}} - \frac{1}{C_\textrm{geo} + \delta c} \right) \approx \frac{\partial\mu}{\partial n} + \frac{Ae^2\delta c}{C_\textrm{geo}^2}.
	\label{error2}
\end{equation}
The change in computed chemical potential across a range of density $\Delta n$ is therefore
\begin{equation}
	\int_{\Delta n} Ae^2\left(\frac{1}{C_\textrm{T}} - \frac{1}{C_\textrm{geo} + \delta c} \right)dn = \Delta \mu + \frac{Ae^2\delta c}{C_\textrm{geo}^2}\Delta n
	\label{error3}
\end{equation}
where $\Delta \mu$ represents the true change in chemical potential and $\frac{Ae^2\delta c}{C_\textrm{geo}^2}\Delta n$ the associated systematic error. If we use the value $\delta c = \SI{0.014}{\fF}$ based on our calculated estimate of $C_\textrm{geo}$ and a visual analysis of the field-dependent data, the errors associated with the gaps at $n_\textrm{s}/4$ and $n_\textrm{s}/2$ are found to be $\delta\left(\Delta_{n_\textrm{s}/4}\right) =  \SI{1.0}{\milli\electronvolt}$ and $\delta\left(\Delta_{n_\textrm{s}/2}\right) =  \SI{1.2}{\milli\electronvolt}$, respectively. The larger error for $\Delta_{n_\textrm{s}/2}$ is due to its slightly larger span in carrier density $\Delta n$.

\subsection{Capacitance measurements at $\SI{5}{\kelvin}$.}
\begin{figure}[h!tp]
\centering
\includegraphics[scale=1]{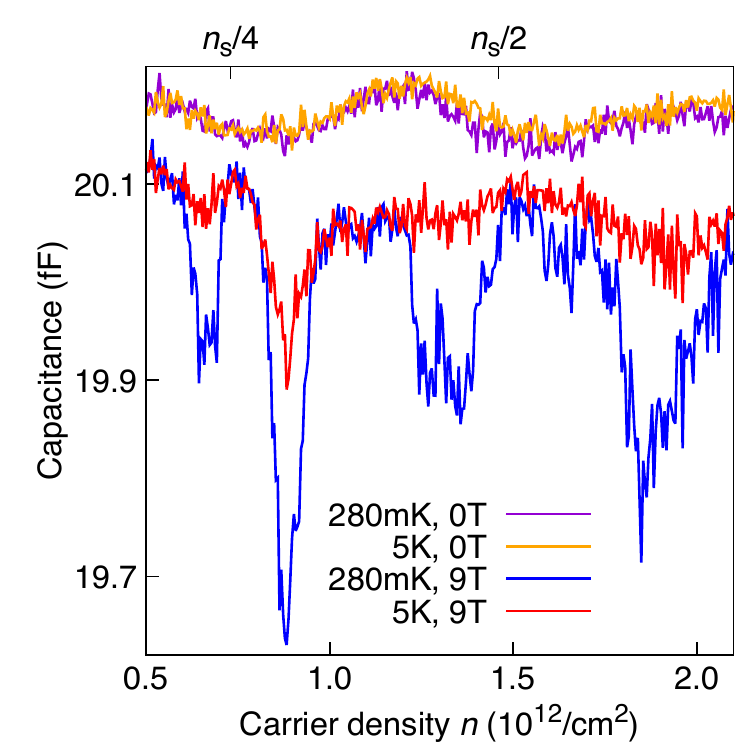}
\caption{Plot of the temperature dependence of the capacitance on the electron-side near commensurate fillings. The purple (orange) trace shows the zero-field capacitance at $\SI{280}{\milli\kelvin}$ ($\SI{5}{\kelvin}$) while the blue (red) trace shows the capacitance at $B=\SI{9}{\tesla}$ at $\SI{280}{\milli\kelvin}$ ($\SI{5}{\kelvin}$). The $\SI{9}{\tesla}$ data has been shifted down for clarity. The capacitance was measured with a $\SI{2.8}{\milli\volt}$ RMS excitation at $\SI{30}{\kilo\hertz}$.}
\label{hightemp}
\end{figure}

Upon warming to $\SI{5}{\kelvin}$ the capacitance of sample M2 shows essentially no change near the commensurate filling gaps on the electron side at zero magnetic field as shown in Fig.~\ref{hightemp}. The purple (orange) trace shows the capacitance at zero magnetic field and $\SI{280}{\milli\kelvin}$ ($\SI{5}{\kelvin}$). The lack of temperature evolution implies that the energy gaps at commensurate filling are well in excess of $\SI{5}{\kelvin}$, consistent with our gap estimation in Fig.~2. At $B=\SI{9}{\tesla}$ the capacitance minima are suppressed upon warming from $\SI{280}{\milli\kelvin}$ (blue) to $\SI{5}{\kelvin}$ (red). At high magnetic field our measurements are no longer in the low frequency limit due to the large in-plane resistivity of the sample while in a quantum Hall gap. The capacitance minima at base temperature are exaggerated by the failure of the sample to charge completely on each excitation cycle. Upon warming to $\SI{5}{\kelvin}$ the in-plane conductivity increases, leading to a reduction in the capacitance features. This evolution at high magnetic field with temperature is likely not related to a change in the thermodynamic density of states, but rather in-plane transport features. 

\subsection{Bandwidth estimation}
\begin{figure}[h!tp]
\centering
\includegraphics[scale=1]{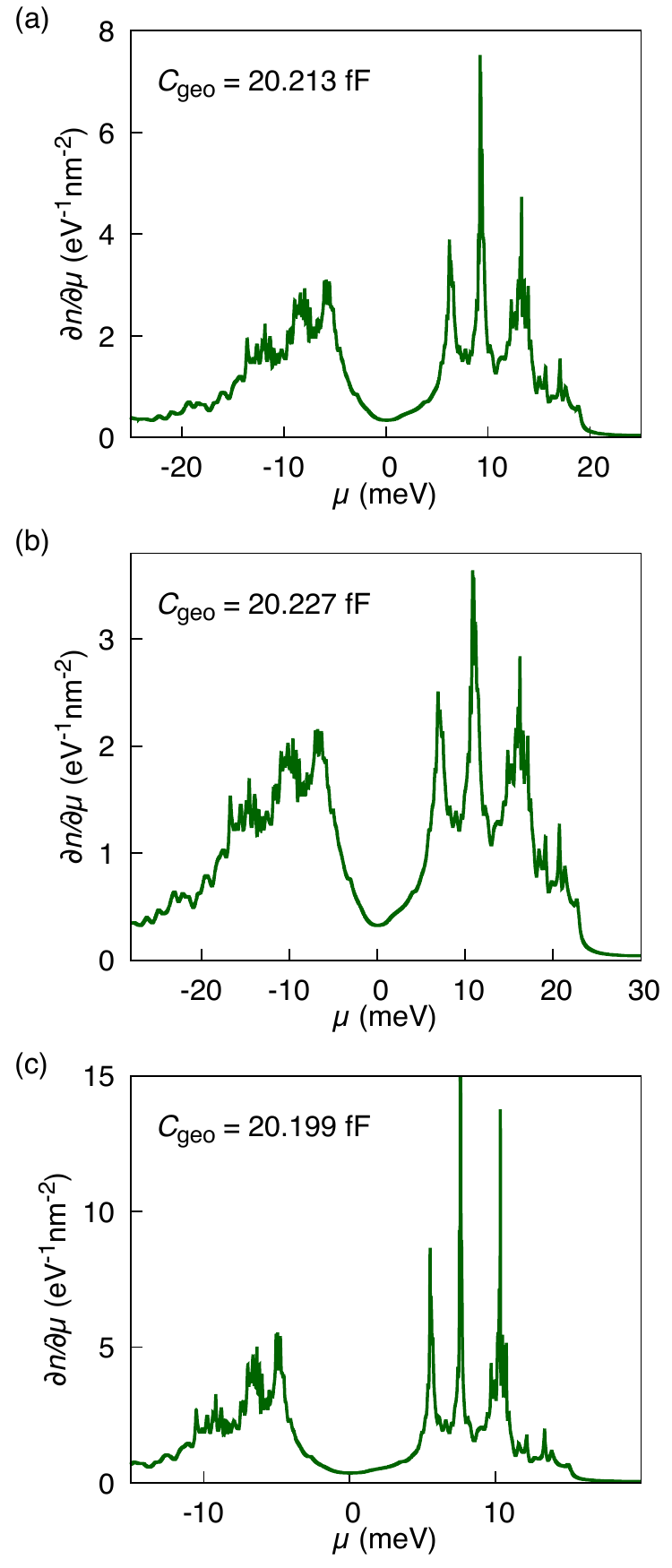}
\caption{\textbf{(a)} Plot of $\partial n/\partial \mu$ for best estimate of $C_\textrm{geo} = \SI{20.213}{\fF}$. \textbf(b) Plot of $\partial n/\partial \mu$ for $C_\textrm{geo} + \delta c = \SI{20.227}{\fF}$. (c) Plot of $\partial n/\partial \mu$ for $C_\textrm{geo} - \delta c = \SI{20.199}{\fF}$. The central density of states peak on the electron side rises to about \SI[per-mode=reciprocal]{55}{\per\electronvolt\per\nano\meter\squared}}
\label{compcgeo}
\end{figure}

When plotting the compressibility in Fig.~3(d), uncertainty in the precise value of $C_\textrm{geo}$ can contribute an uncertainty in the movement of the chemical potential with density. As detailed previously in Eqs.~\ref{error1}--\ref{error3}, the uncertainty in the shift of the chemical potential $\delta(\Delta \mu)$ can be related to the uncertainty in the geometric capacitance $\delta c$ through
\begin{equation}
	\delta(\Delta \mu) = \frac{Ae^2\delta c}{C_\textrm{geo}^2}\Delta n.
 \end{equation}
Because our total bandwidth spans a density approximately given by $\Delta n = \SI{6e12}{\per\centi\meter\squared}$, the associated error in bandwidth is given by $\delta(\Delta \mu) \approx \SI{10}{\milli\eV}$. In Fig.~3(d) we find a bandwidth of approximately $\SI{35}{\milli\eV}$. Incorporating our estimated uncertainty, the bandwidth has a range that spans approximately $25-\SI{45}{\milli\eV}$. In Fig.~\ref{compcgeo} we plot the compressibility for our best estimate of $C_\textrm{geo}$ as well as the upper and lower bounds of our uncertainty estimate $C_\textrm{geo} \pm \delta c$ for $\delta c = \SI{0.014}{\fF}$. The bandwidth range is roughly $25-\SI{45}{\milli\eV}$ in line with our error calculation. The vertical axis is very sensitive to the specific choice of $C_\textrm{geo}$ whenever $C_\textrm{T} \approx C_\textrm{geo}$ which is why the highly compressible peaks appear at such different values of $\partial n/\partial \mu$. The lower compressibility features (e.g. charge neutrality, commensurate filling on the electron-side) show much less variation. The plot in Fig.~\ref{compcgeo}(c) has been cut off above $\SI[per-mode=reciprocal]{15}{\per\electronvolt\per\nano\meter\squared }$, where the central density of states maximum rises to about $\SI[per-mode=reciprocal]{55}{\per\electronvolt\per\nano\meter\squared }$, in order to more easily compare the low-compressibility features between panels.

\subsection{Loss tangent in finite magnetic field}
\begin{figure*}[h!tp]
\centering
\includegraphics[scale=1]{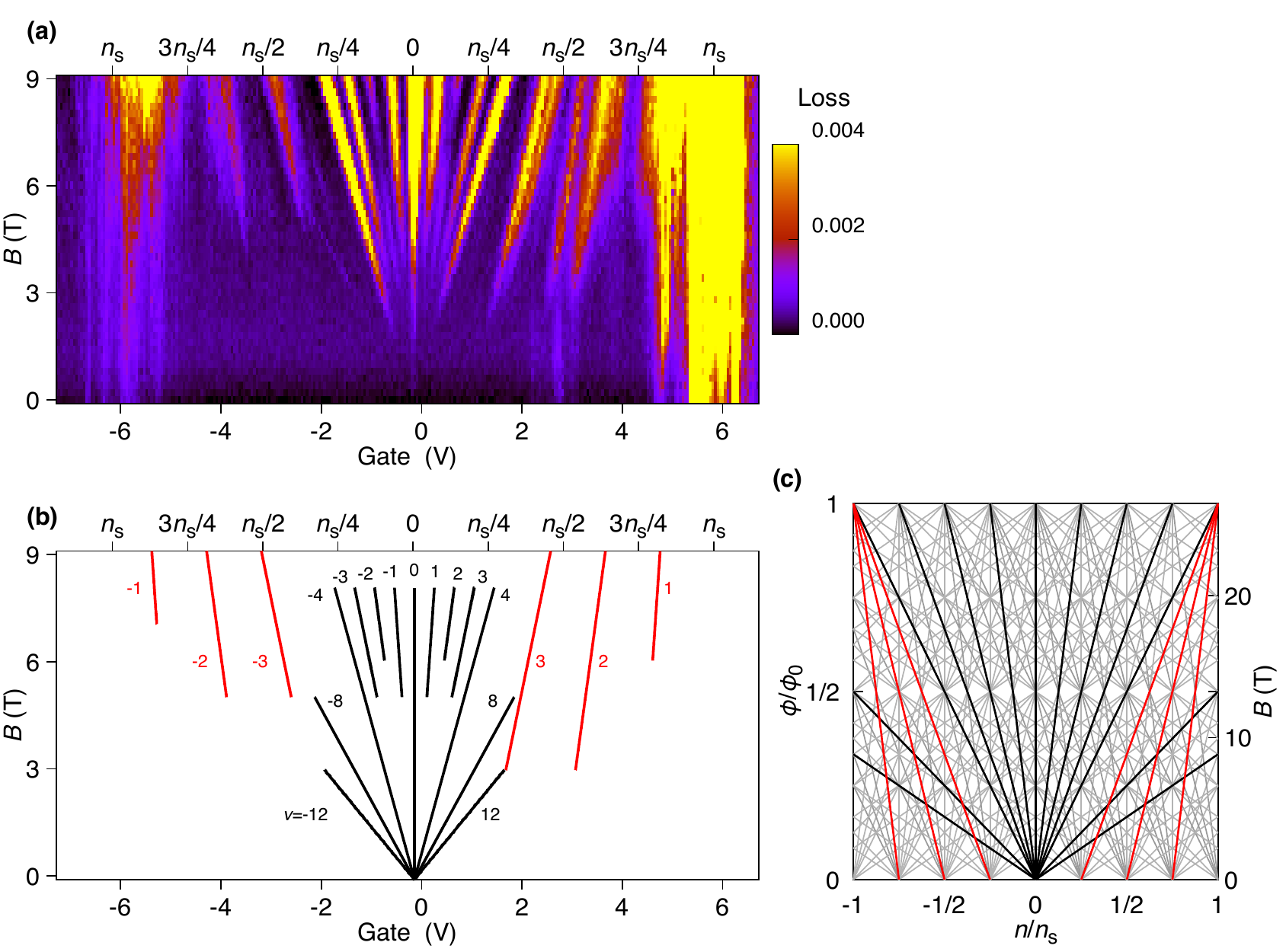}
\caption{\textbf{(a)} Plot of the loss tangent of device M2 as a function of carrier density and magnetic field. Weak resistive features around $n_\textrm{s}/2$ are visible and track vertically with magnetic field until being obscured by the fractal miniband gaps around $\SI{3}{\tesla}$. Multiple vertical features adjacent to one another indicate that the twist angle may be inhomogeneous through the entire sample. The color scale has been suppressed above $0.004$ in order to reveal weaker features at low magnetic field. The loss tangent was measured with a $\SI{2.8}{\milli\volt}$ RMS excitation at $\SI{150}{\kilo\hertz}$. \textbf{(b)} Schematic from Fig.~4(b) tracking the filling factors of the field-induced gaps. \textbf{(c)} Wannier diagram of associated field-induced gaps observed in (b).}
\label{lossm2}
\end{figure*}

In Fig.~\ref{lossm2}(a) we plot the loss tangent of device M2 as a function of carrier density and magnetic field in order to reveal additional information about the in-plane conductivity of the sample. The loss tangent is given by $\omega R_\textrm{in-plane}C_\textrm{T}$. Increases in the loss tangent correspond to increases in the in-plane resistance (which tend to dominate any changes in $C_\textrm{T}$ at high magnetic field for this device) as a resistive state is entered and serve as a qualitative measure of the in-plane bulk transport. The bright features emanating from charge neutrality are the cyclotron and exchange gaps arising from the quantum Hall regime whereas the bright features that emanate from high magnetic field and terminate near the commensurate fillings arise from fractal ``Hofstadter'' minibands due to the interaction of the magnetic field and superlattice potential. Importantly, the half-filling state on the electron doped regime shows faint vertical features which appear to terminate around $\SI{3}{\tesla}$, though they becomes partially obscured due to the coexistence of the fractal miniband feature. This indicates that the resistive features survive to at least $\SI{3}{\tesla}$. There are multiple closely spaced resistive features which are grouped around the half-filling location indicating possible inhomogeneity in the rotation angle across the lateral extent of the sample. Importantly, we do not see doubling of the central Landau fan indicating that the charge density across the sample remains uniform. We do not see noticeable resistive features associated with either the hole states or the quarter-filled electronic state. In panel (b) of Fig.~\ref{lossm2} we plot the same filling factor schematic as in Fig.~4(b). The Wannier diagram in panel (c) of Fig.~\ref{lossm2} shows the possible fractal miniband gaps in grey with associated observed gaps in (a) color coded to match (b).

\subsection{Magnetic field dependence of device M1}
\begin{figure*}[h!tp]
\centering
\includegraphics[scale=1]{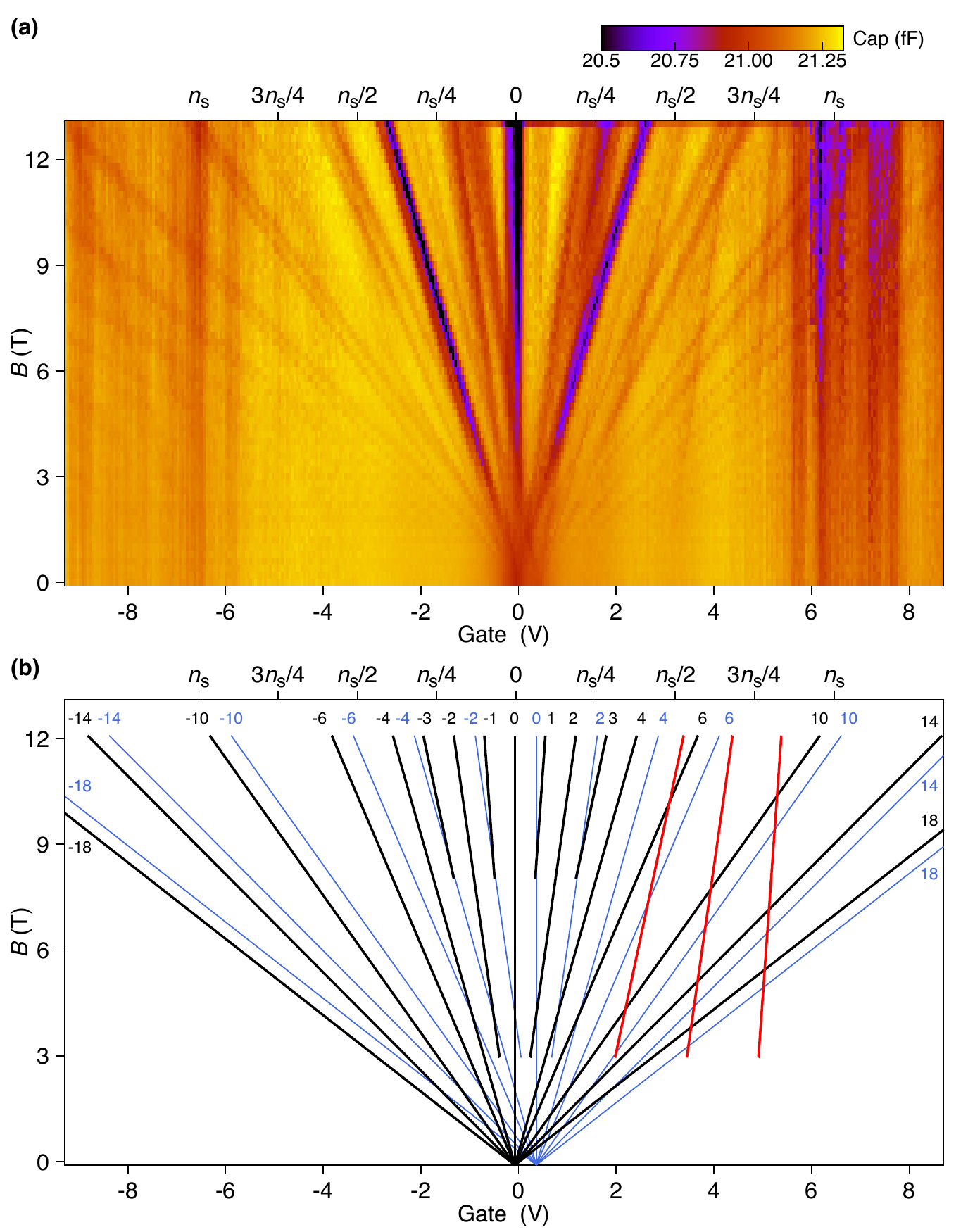}
\caption{\textbf{(a)} Magnetic field dependence of device M1. The color scale has been suppressed below $\SI{20.5}{\fF}$. \textbf{(b)} Schematic showing some of the important incompressible phases of device M1. The black lines indicate cyclotron or exchange gaps arising from the central Landau fan. The blue lines indicate the gaps arising from an additional, weaker Landau fan, indicating device M1 contains a second region at slightly different doping. Filling factors for both fans are labeled. The red lines indicate field-induced gaps which terminate at the commensurate filling associated with fractal miniband gaps as discussed in the main text in Fig.~4. In contrast to device M2, the fractal miniband gaps do not appear doubled, indicating improved twist angle uniformity. The capacitance was measured with a $\SI{2.8}{\milli\volt}$ RMS excitation at $\SI{150}{\kilo\hertz}$ at $\SI{225}{\milli\kelvin}$.}
\label{fieldm1}
\end{figure*}

In Fig.~\ref{fieldm1} we plot the magnetic field dependence of device M1. Adjacent to the main Landau fan is a second weaker fan emanating from a displaced Dirac point, indicating that the sample contains a second region which is at a slightly different doping. This may be associated with a region of the device adjacent to one of the ohmic contacts away from the central portion of the etched Hall bar geometry. Similar incompressible phases (red lines in Fig.~\ref{fieldm1}(b)) emerge from high magnetic field on the electron-doped side and tend towards commensurate filling locations on the abscissa as discussed in the main text in Fig.~4(a) for device M2. Here, the gaps emanating from high field do not appear doubled as in Fig.~4 indicating improved homogeneity in the rotation angle.  

\end{document}